\documentclass{elsart5p}

\usepackage{amssymb}
\usepackage{amsmath}
\usepackage{graphicx}

\begin{document}

\begin{frontmatter}


\title{The $^4$He total photo-absorption cross section with two- plus three-nucleon interactions from chiral effective field theory}


\author{Sofia Quaglioni},
\ead{quaglioni1@llnl.gov}
\author{Petr Navr\'atil}
\ead{navratil1@llnl.gov}
\address{Lawrence Livermore National Laboratory, L-414, P.O. Box 808, Livermore, CA 94551, USA}


\begin{abstract}
The total photo-absorption cross section of $^4$He is evaluated microscopically 
using two- (NN) and three-nucleon (NNN) interactions based upon chiral effective 
field theory ($\chi$EFT). The calculation is performed using the Lorentz integral 
transform method along with the {\em ab initio} no-core shell model approach.
An important feature of the present study is the consistency of the NN and NNN
interactions and also, through the Siegert theorem, of the two- and three-body current 
operators. This is due to the application of the $\chi$EFT framework. 
The inclusion of the NNN interaction produces a suppression of the low-energy peak 
 and enhancement of the high-energy tail of the cross section.
We compare to calculations obtained using other interactions and to representative experiments.
The rather confused experimental situation in the giant resonance region prevents 
discrimination among different interaction models.
\end{abstract}

\end{frontmatter}

Interactions among nucleons are governed by quantum chromodynamics (QCD). 
In the low-energy regime relevant to nuclear structure and reactions, 
this theory is non-perturbative, and, therefore, hard to solve. Thus, theory has 
been forced to resort to models for the interaction, which have limited physical basis. 
New theoretical developments, however, allow us to connect QCD 
with low-energy nuclear physics. Chiral effective field theory 
($\chi$EFT)~\cite{Weinberg:1979,Weinberg:90-91} 
provides a promising bridge to the underlying theory, QCD. 
Beginning with the pionic or the nucleon-pion system~\cite{bernard95} 
one works consistently with systems of increasing 
number of nucleons~\cite{ORK94-Bira-bedaque02a}. 
One makes use of spontaneous breaking of chiral symmetry to systematically 
expand the strong interaction in terms of a generic small momentum
and takes the explicit breaking of chiral symmetry into account by expanding 
in the pion mass. 
Nuclear interactions are non-perturbative, because diagrams with purely nucleonic
intermediate states are enhanced \cite{Weinberg:1979,Weinberg:90-91}. Therefore, the chiral perturbation 
expansion is performed for the potential. 
The $\chi$EFT predicts, along with the nucleon-nucleon (NN) interaction at the leading order, 
a three-nucleon (NNN) interaction at the next-to-next-to-leading order or 
N$^2$LO~\cite{Weinberg:90-91,VanKolck:1994,Epelbaum:2002}, 
and even a four-nucleon (NNNN) interaction at the fourth order (N$^3$LO)~\cite{Epelbaum06}.
The details of QCD dynamics are contained in parameters, 
low-energy constants (LECs), not fixed by the symmetry, but can be constrained 
by experiment. At present, high-quality NN potentials have been determined 
at N$^3$LO~\cite{Entem:2003}. 
A crucial feature of $\chi$EFT is the consistency between 
the NN, NNN and NNNN parts. As a consequence, at N$^2$LO and N$^3$LO, except 
for two parameters assigned to two NNN diagrams, the potential is fully 
constrained by the parameters defining the NN interaction. The full interaction up to N$^2$LO
was first applied to the analysis of $nd$ scattering~\cite{Epelbaum:2002} and later the N$^3$LO NN potential was combined with the available NNN at N$^2$LO to study the $^7$Li structure~\cite{nogga:064002}.
In a recent work~\cite{Navratil:2007} the NN potential 
at N$^3$LO of Ref.~\cite{Entem:2003} and the NNN interaction at 
N$^2$LO~\cite{VanKolck:1994,Epelbaum:2002} have been applied to the calculation 
of various properties of $s$- and mid-$p$-shell nuclei, using the {\em ab initio} 
no-core shell model (NCSM)~\cite{Navratil:2000wwgs,Navratil:152502}, 
up to now the only approach able to handle the chiral NN+NNN potentials 
for systems beyond $A=4$. 
In that study, a preferred choice of the two NNN LECs was found 
and the fundamental importance of the chiral NNN interaction was demonstrated 
for reproducing the structure of light nuclei. In the present work, we apply for 
the first time the
same $\chi$EFT interactions to the {\em ab inito} calculation of reaction observables 
involving the continuum of the four-nucleon system. In particular, we study the $^4$He 
total photo-absorption cross section.


Experimental measurements of the $\alpha$ particle photo-disintegration suffer from a recurrent history of large discrepancies in the near-threshold region, where the $^4$He$(\gamma,p)^3$H and the  $^4$He$(\gamma,n)^3$He break-up channels dominate the total photo-absorption cross section (we refer the reader to the reviews of available data in Refs.~\cite{CBD:1983,quaglioni:044002,Shima:2005}).  The latest examples date back to the past two years~\cite{Shima:2005,Nilsson:2005}. Of particular controversy is the height of the cross section at the peak, alternatively found to be either pronounced or suppressed with differences up to a factor of 2 between different experimental data. 
With the exception of~\cite{Ellerkmann:1996}, early evaluations of the $^4$He photo-disintegration~\cite{ELO:1997B,BELO:2001,quaglioni:044002} showed better agreement with the high-peaked experiments, and, ultimately, with those of Ref.~\cite{Nilsson:2005}. The inability of these calculations to reproduce a suppressed cross section at low energy was often imputed to the semi-realistic nature of the Hamiltonian and, in particular, to the absence of the NNN force. The introduction of NNN interactions leads, indeed, to a reduction of the peak height, as it was recently shown in a calculation of the photo-absorption cross section with the Argonne V18 (AV18) NN potential augmented by the Urbana IX (UIX) NNN force~\cite{Gazit:112301}.
A damping of the peak was also found using the correlated AV18 potential constructed within the unitary correlation operator method (UCOM)~\cite{Bacca:2007}. In both cases, however, the suppression is not sufficient to reach the low-lying data, and in particular those of Ref.~\cite{Shima:2005}.
The latter calculations  represent a substantial step forward in the study of the $^4$He photo-disintegration. 
However,  they still present a residual degree of arbitrariness in the choice of the NNN force to complement AV18 in the first case, or in the choice of the unitary transformation leading to the non-local phase-equivalent interaction in the second case.
We note that  the Illinois potential models have been found to be more realistic NNN partners of AV18 in the reproduction of the structure of light $p$-shell nuclei~\cite{PhysRevC.64.014001}.  From a fundamental point of view, it is therefore important to calculate the $^4$He photo-absorption cross section in the framework of $\chi$EFT theory, where NN and NNN potentials are derived in a consistent way and their relative strengths is well established by the order in the chiral expansion.

When the wavelength of the incident radiation is much larger than the spatial extension of the system under consideration,
the nuclear photo-absorption process can be described in good approximation by the cross section 
\begin{equation}
\label{sigma}
\sigma_\gamma(\omega)=4\pi^2\frac{e^2}{\hbar c}\omega R(\omega)\;,
\end{equation}
where $\omega$ is the incident photon energy and  the inclusive response function
\begin{equation}
\label{response}
R(\omega)=\int d\Psi_{f}\left|\left\langle \Psi_{f}\right|\hat D\left|\Psi_{0}\right\rangle \right|^{2}\delta(E_{f}-E_{0}-\omega)\,
\end{equation}
is the sum of all  the transitions from the ground state $|\Psi_0\rangle$ to the various allowed final states $|\Psi_f\rangle$ induced by the dipole operator:
\begin{equation}
\label{dipole}
\hat D=\sqrt{\frac{4\pi}{3}}\sum_{i=1}^A \frac{\tau^z_i}{2} r_i  Y_{10}(\hat{r}_i)\;.
\end{equation}
In the above equations ground- and final-state energies are denoted by $E_0$ and $E_f$, respectively, whereas $\tau^z_i$ and $\vec{r\,}_i=r_i\hat{r}_i$ represent the isospin third component and center of mass frame coordinate of the {\em i}th nucleon.
This form of the transition operator includes the leading effects of the meson-exchange currents through the Siegert's theorem.  
Additional contributions to the cross section (due to retardation, higher electric multiples, magnetic multiples) not considered by this approximation are found to be negligible in the $A=2$~\cite{AS:1991} and $A=3$~\cite{Golak:2002} nuclei, in particular for $\omega\lesssim 40$ MeV.  A similar behavior can be expected from a system of small dimensions like the $^4$He.

Denoting by $\hat H$ the full Hamiltonian of the system, 
\begin{equation}
\hat H = \frac{1}{A}\sum_{i<j}^A\frac{\vec{p\,}_i-\vec{p\,}_j}{2m} +\sum_{i<j}^A V^{\rm NN}_{ij}+\sum_{i<j<k}^A V^{\rm NNN}_{ijk}\;,
\label{hbare}
\end{equation}
where $m$ is the nucleon mass, $V^{NN}_{ij}$ is the sum of N$^3$LO NN and Coulomb interactions, and  $V^{NNN}_{ij}$ is the N$^2$LO NNN force, we $i$) solve the many-body Schr{\"o}dinger equation for the ground state $|\Psi_0\rangle$, $ii$) obtain the  response~(\ref{response}) by evaluation~\cite{EFROS:859399,MBLO:2003} and subsequent inversion~\cite{ELO:1999-Andreasi:2005} of an integral transform with a Lorentzian kernel of finite width $\sigma_I\sim\,$10$-$20 MeV ($z=E_0+\sigma_R+i\sigma_I$),
\begin{eqnarray}
\label{lit1}
L(\sigma_{R},\sigma_{I})&=&-\frac{1}{\sigma_I}{\rm Im}\left\{\langle\psi_0|\hat D^\dagger \frac{1}{z-\hat H}\hat D|\psi_0\rangle\right\}\,\\
\label{lit2}
&=& \int R(\omega)\,\frac{1}{(\omega-\sigma_{R})^{2}+\sigma_{I}^{2}}\, d\omega\;,
\end{eqnarray}
and $iii$) calculate the photo-absorption cross section in the long wave-length approximation using Eq.~(\ref{sigma}). 
Following these steps, a fully microscopic result for the $^4$He photo-absorption cross section  can be reached  through the use of efficient expansions over localized many-body states. Indeed, in the technique summarized by Eqs.~(\ref{lit1}-\ref{lit2}) and known as Lorentz integral transform (LIT) method~\cite{Efros:1994iq}, the continuum problem is mapped onto a bound-state-like problem. 

The present calculations are performed in the framework of  the {\em ab initio} NCSM approach~\cite{Navratil:2000wwgs}.  This method looks for the eigenvectors of $\hat H$ in the form of  expansions over a complete set of harmonic oscillator (HO) basis states up to a maximum excitation of $N_{max}\hbar\Omega$ above the minimum energy configuration, where $\Omega$ is the HO parameter. 
The convergence to the exact results with increasing $N_{max}$ is accelerated by the use of  an effective interaction derived, in this case,  from the adopted NN and NNN $\chi$EFT potentials at the three-body cluster level~\cite{Navratil:152502}.  The reliability of the NCSM approach combined with the LIT method was validated  by comparing to the results obtained with the effective-interaction hyper-spherical harmonics (EIHH) technique~\cite{barnea:054001-03} in a recent benchmark calculation~\cite{Stetcu:2007}. 
A complete description of the NCSM approach was presented, e.g., in Refs.~~\cite{Navratil:2000wwgs,Navratil:152502}. Here, we emphasize some of the aspects involved in a calculation of the effective interaction at the three-body cluster level in presence of a NNN potential. We use the Jacobi coordinate
HO basis antisymmentrized according to the method described in Ref.~\cite{Navratil:1999pw}.
The NCSM calculation proceeds as follows. First, we diagonalize the Hamiltonian 
with and without the NNN interaction in a three-nucleon basis for all relevant three-body channels.
Second, we use the three-body solutions from the first step to derive three-body 
effective interactions with and without the NNN interaction. By subtracting the two effective interactions
we isolate the NN and NNN contributions. This is needed due to a different scaling with particle number of the two- and the three-body interactions. 
The $^4$He effective interaction is then obtained by adding the two contributions 
with the appropriate scaling factors~\cite{Navratil:152502}. 
Note that our effective interaction is model-space dependent. Consequently, we need both the effective interaction for the $^4$He ground state ($J^\pi T=0^+ 0$), and the one for the $1^- 1$ states, entering the LIT calculation. Indeed, due to the change of parity, the model-space size changes ($N_{\rm max}\rightarrow N_{\rm max}+1$). With the effective interactions replacing the interactions
in the Hamiltonian~(\ref{hbare}), the four-nucleon calculations proceed as described in the text following Eq.~(\ref{hbare}). 

We start our discussion presenting the results obtained for the ground state of the $\alpha$ particle using two different values of the HO parameter, namely $\hbar\Omega=22$ and $28$ MeV. 
This choice for the HO frequencies is driven by our final goal of evaluating the $^4$He photo-absorption cross section and providing an estimate for its theoretical uncertainty. Indeed, in the particular case of the $^4$He nucleus, frequencies in the range $12\le\hbar\Omega\le28$ MeV allow to achieve a good description of both ground state and  complex energy continuum, as required in a calculation of response functions with the LIT method~\cite{Stetcu:2007}.

\begin{figure}[t]
\includegraphics*[scale=0.62]{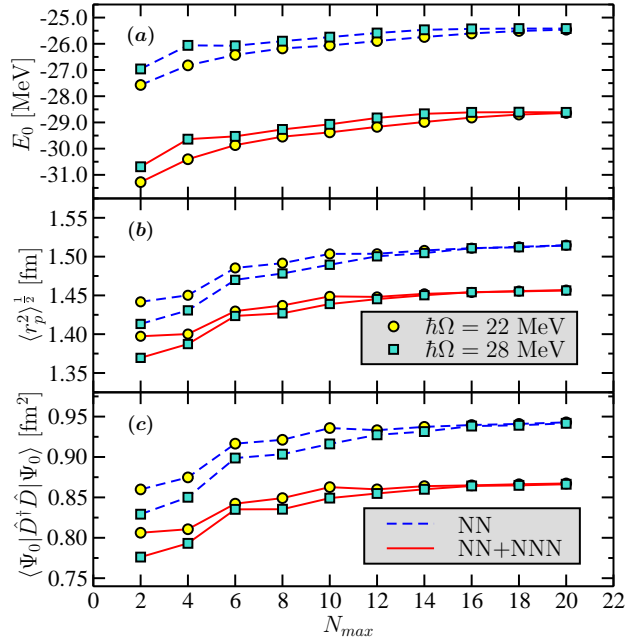}%
\caption{(Color online) The $^4$He ground-state energy $E_0$ [panel $a$)], point-proton root-mean-square  radius  $\langle r_p^2\rangle^{\frac12}$ [panel $b$)] and total dipole strength $\langle\Psi_0|\hat D^\dagger\hat D|\Psi_0\rangle$ [panel $c$)] obtained with the $\chi$EFT NN and NN+NNN interactions. Convergence pattern with respect to the model space truncation $N_{max}$ for $\hbar\Omega=22$ and $\hbar\Omega=28$ MeV.\label{gsfig}}
\end{figure}
%
For all of the three observables 
examined in Fig.~\ref{gsfig}
the $\chi$EFT NN and NN+NNN interactions lead to very similar and smooth convergence patterns.
 In particular, an accurate convergence is reached starting from $N_{max}=18$, as we find independence from both model space and frequency. Although $\chi$EFT forces are known to present a relatively soft core, the use of effective interactions for both the NN and NNN forces is the essential key to this remarkable result. The summary of the extrapolated ground-sate properties is presented in Table~\ref{gstab}. 
 \begin{table}[b]
\caption{Calculated $^4$He ground-state energy $E_0$,  point-proton  root-mean-square radius $\langle r_p^2\rangle^{\frac12}$, and total dipole strength  $\langle\Psi_0|\hat D^\dagger\hat D|\Psi_0\rangle$ obtained using the $\chi$EFT NN and NN+NNN interactions compared to experiment. The experimental value of the  point-proton radius is deduced from the measured alpha-particle charge radius, $\langle r_c^2\rangle^{\frac12}=1.673(1)$ fm~\cite{Borie:1978}, proton charge radius,   $\langle R_p^2\rangle^{\frac12}=0.895(18)$ fm~\cite{Sick:2003}, and neutron mean-square-charge radius, $\langle R_n^2\rangle=-0.120(5)$ fm$^2$~\cite{PhysRevLett.74.2427}. }
\label{gstab}
\newcommand{\m}{\hphantom{$-$}}
\newcommand{\cc}[1]{\multicolumn{1}{c}{#1}}
\renewcommand{\tabcolsep}{0.8pc} 
\renewcommand{\arraystretch}{1.5} 
\begin{tabular}{@{}lcccc}
\hline
&$E_0$ [MeV]&$\langle r_p^2\rangle^{\frac{1}{2}}$ [fm]&$\langle\Psi_0|\hat D^\dagger\hat D|\Psi_0\rangle$ [fm$^2$]\\
\hline
NN & -25.39(1) & 1.515(2) & 0.943(1)\\
NN+NNN & -28.60(3)& 1.458(2) & 0.868(1)\\
Expt. & -28.296\phantom{()} & 1.455(7) & -\\
\hline
\end{tabular}
\\[2pt]
\end{table}
%
The present results for ground-state energy and point-proton radius with the N$^3$LO NN interaction are consistent with a previous NCSM evaluation ($E_0=-25.36(4)$ MeV, $\langle r_p^2\rangle^{\frac{1}{2}}=1.515(10)$ fm) obtained using a two-body effective interaction in a model space up to $N_{max}=18$~\cite{Navratil:014311} and with that obtained by the hyper-spherical harmonic variational calculation of Ref.~\cite{Viviani:2006} ($E_0=-25.38$ MeV, $\langle r_p^2\rangle^{\frac{1}{2}}=1.516$ fm ) and by the Faddeev-Yakubovsky method~\cite{NoggaFY} ($E_0=-25.37$ MeV). Finally, with the present choice for the LECs~\cite{Navratil:2007} the calculated binding-energy with inclusion of the NNN force is within few hundred KeV of experiment. This leaves room for additional effects expected from the inclusion of the here missing N$^3$LO NNN (not yet available) and NNNN interaction terms~\cite{Rozdpedzik:2006}. 

At the ground-state level, the inclusion of the NNN force affects mostly the energy, providing 3.21 MeV additional binding, while only a weak suppression of about 3.8\% is found for the point-proton radius. That the  total dipole strength follows the same pattern as the radius and is reduced of 7.9\% is not so surprising considering the approximate relation between  them~\cite{Foldy:1957}:
\begin{equation}
\label{sumrule1}
\langle\Psi_0|\hat D^\dagger\hat D|\Psi_0\rangle \simeq 
\frac{ZN}{3(A-1)}\langle r_p^2\rangle\,.
\end{equation}
The latter expression, which is exact for the deuteron and the triton and for ground-state wave functions symmetric under exchange of the spatial coordinates of any pair of nucleons, represents a quite reasonable approximation for the $\alpha$-particle and is found to be within 9\% off our calculations with both the NN and NN+NNN $\chi$EFT potentials. As we will see later, this also implies rather weak  NNN effects on the $^4$He photo-absorption cross section at low energy.  

\begin{figure}[t]
\includegraphics*[scale=0.62]{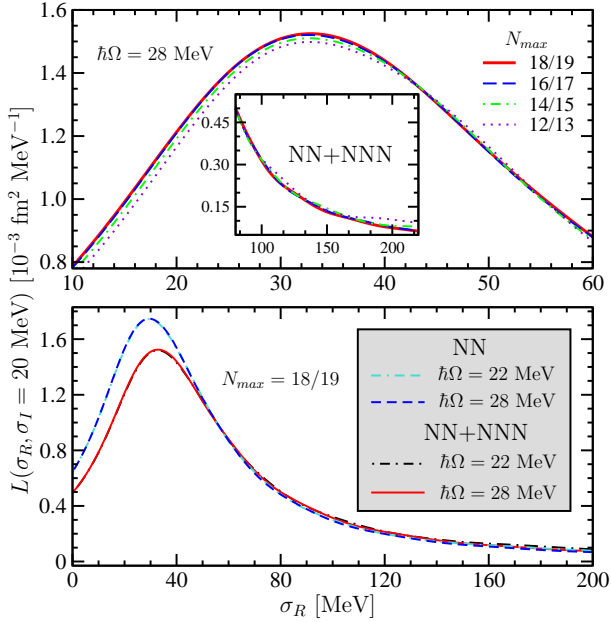}%
\caption{(Color online) The LIT of the $^4$He dipole response as a function of $\sigma_R$ at $\sigma_I=20$ MeV. Convergence pattern of the NN+NNN calculation with respect to the model-space truncation $N_{max}$ for $\hbar\Omega=28$ MeV  (upper panel),  and frequency dependence of the best ($N_{max}=18/19$) results with and without inclusion of the NNN force (lower panel).  \label{litfig}}
\end{figure}

We turn now to the second part of our calculation, for which the ground state is an input. The actual evaluation of Eq.~(\ref{lit1}) is performed by applying the Lanczos algorithm to the Hamiltonian of the system, using as starting vector $|\varphi_0\rangle=\langle\Psi_0|\hat D^\dagger\hat D|\Psi_0\rangle^{-\frac{1}{2}}\hat D|\Psi_0\rangle$~\cite{MBLO:2003,Stetcu:2007}. Indeed, the LIT can be written as a continued fraction of the elements of the resulting tridiagonal matrix, the so-called Lanczos coefficients $a_n$ and $b_n$:
\begin{equation}
\label{lit3}
L(\sigma)=\frac{1}{\sigma_I}{\rm Im}\frac{\langle\Psi_0|\hat D^\dagger\hat D|\Psi_0\rangle}{(z-a_0)-\frac{b_i^2}{(z-a_1)-\frac{b_2^2}{(z-a_2)-\frac{b_3^2}{\cdots}}}}\;.
\end{equation}  
Due to the selection rules induced by the dipole operator~(\ref{dipole}), for a given truncation $N_{max}$ in the $0^{+}0$ model space used to expand the ground state,  a complete calculation of Eq.~(\ref{lit3}) requires an expansion of $|\varphi_0\rangle$ over a $1^{-}1$ space up to $N_{max}+1$. This is the oriigin of the even/odd notation for $N_{max}$ introduced to describe the convergence of the  LIT in Fig.~\ref{litfig}.
The LITs obtained using the NN and NN+NNN $\chi$EFT interactions show, once again, convergence patterns very similar to each other.
As an example, in the upper panel of Fig.~(\ref{litfig}) we show the model-space dependence of the LIT including the NNN force at  $\hbar\Omega=28$ MeV. Thanks to the use of three-body effective interaction for both the NN and NNN terms of the potential, a stable position and height of the peak  in the low-$\sigma_R$ region and satisfactory quenching of the oscillations in the tail  are found for  $N_{max}=18/19$. In this regard, our approach differs from the one of Ref.~\cite{Gazit:112301}, where the effective interaction (at the two-body cluster level) is constructed only for the NN potential, while the NNN force is taken into account as bare interaction. The bottom panel of Fig.~\ref{litfig}  indicates that for $N_{max}=18/19$ we find also a fairly good  agreement between the $\hbar\Omega=22$ and $\hbar\Omega=28$ MeV calculations, in particular below $\sigma_R=60$ MeV, where for both NN and NN+NNN interactions the two curves are within 0.5\% of each other.  At higher $\sigma_R$ the $\hbar\Omega=22$ MeV results present a weak oscillation (less than 5\% in the range $60\leq\sigma_R\leq140$ MeV) around the $\hbar\Omega=28$ MeV curves, and the discrepancy between the two frequencies becomes larger beyond $\sigma_R=140$ MeV, where the absolute value of the LIT is small. As we will see later, this small discrepancy will be propagated to the cross section by the inversion procedure~\cite{ELO:1999-Andreasi:2005}, giving rise to the uncertainty of our calculations. As for the NNN effects at the level of the LIT, the shift of about 3 MeV in the position of the peak is due to the different ground-state energies for the NN and NN+NNN potentials. In addition one can notice a quenching of about 13\% of the peak height.  
\begin{figure}[t]
\includegraphics*[scale=0.62]{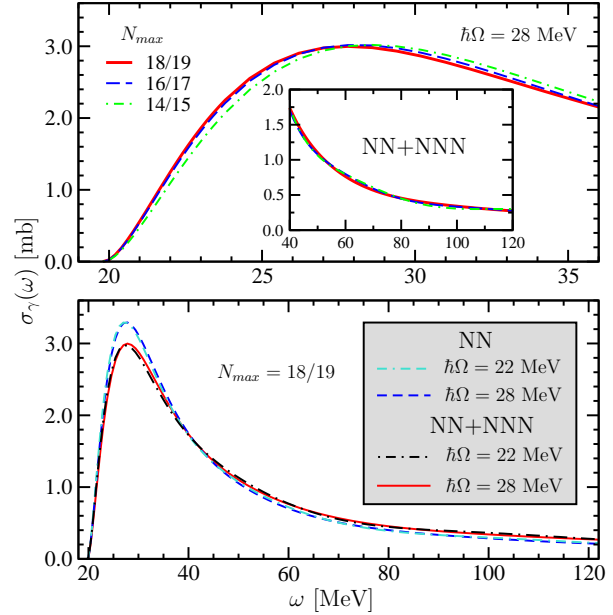}%
\caption{(Color online) The $^4$He photo-absorption cross section as a function of the excitation energy $\omega$. Convergence pattern of the NN+NNN calculation with respect to the model-space truncation $N_{max}$ for $\hbar\Omega=28$ MeV  (upper panel),  and frequency dependence of the best ($N_{max}=18/19$) results with and without inclusion of the NNN force (lower panel). \label{csfig}}
\end{figure}

In analogy with Fig.~\ref{litfig}, Fig.~\ref{csfig} shows the convergence behavior of our results for the cross section. Starting from $N_{max}=14/15$ the calculated LIT's are accurate enough to find stable inversions for the response function, and hence deriving the corresponding results 
for the cross section. 
The curves obtained  for the NN+NNN interaction at the HO frequency value of $\hbar\Omega=28$ MeV are shown in the upper panel: the model space dependence is weak and the difference between $N_{max}=16/17$ and $18/19$ never exceeds 5\% in the range from threshold to $\omega=120$ MeV. A somewhat larger discrepancy (less than 7\%) is  found by comparing the best results ($N_{max}=18/19$) for $\hbar\Omega=22$ MeV and $28$ MeV. As with the LIT, the first oscillates slightly around the second, particularly in the tail of the cross section. We will use this discrepancy as an estimate for the theoretical uncertainty of our calculations. Note that both the NN and NN+NNN calculated cross sections are translated to the experimental threshold for the $^4$He photo-disintegration, $E_{th}=19.8$ MeV ($\omega\rightarrow\omega+\Delta E_{th}$, with $\Delta E_{th}$ being the difference of the calculated and experimental thresholds). The same procedure will be applied later in the comparison with experimental data and different potential models. Under this arrangement, the position of the peak is not affected by the inclusion of the NNN force, while the relative difference between the NN and NN+NNN  cross sections varies almost linearly from $-10\%$ at threshold to about $+25\%$ at $\omega=120$ MeV. In particular, the peak height undergoes a 9\% suppression and the two curves cross around $\omega=40$ MeV.  
In view of the inverse-energy-weighted integral of the cross-section~(\ref{sigma}),
\begin{equation}
\label{sumrule2}
\int_{E_{th}}^\infty \frac{\sigma_{\gamma}(\omega)}{\omega} d\omega = 4\pi^2\frac{e^2}{\hbar c} \langle \Psi_0|\hat D^\dagger\hat D|\Psi_0\rangle\;,
\end{equation}
the mildness of the NNN force effects in the peak region is a consequence of the small reduction found for the total dipole strength. Considering in addition the approximate relation~(\ref{sumrule1}), 
we can infer a weak sensitivity of the cross section at low energy with respect to variations of the LECs in the NNN force, for which we have embraced the preferred choice suggested in Ref.~\cite{Navratil:2007}.

\begin{figure}[t]
\includegraphics*[scale=0.62]{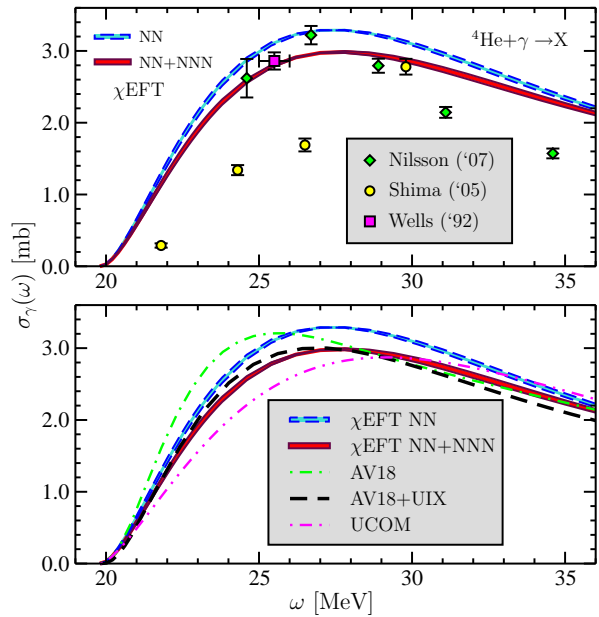}%
\caption{(Color online) The $^4$He photo-absorption cross section as a function of the excitation energy $\omega$. Present NCSM results obtained using the $\chi$EFT NN and NN+NNN interactions compared to:  (upper panel) the $^4$He$(\gamma,n)$ data of Nilsson {\it et al}.~\cite{Nilsson:2005} multiplied by a factor of 2, the total cross section measurements of Shima  {\it et al}.~\cite{Shima:2005}, the total photo-absorption at the peak derived from Compton scattering via dispersion relations from Wells  {\em et al}.~\cite{Wells:1992}; (lower panel) the EIHH predictions for AV18, AV18+UIX~\cite{Gazit:112301} and UCOM~\cite{Bacca:2007}. The widths of the $\chi$EFT NN and $\chi$EFT NN+NNN curves reflect the uncertainties in the calculations (see text).\label{csfig2}} 
\end{figure}

We compare to experimental data in the region $\omega<40$ MeV, where corrections to the unretarded dipole approximation are expected to be largely negligible and the relative uncertainty of our calculations is minimal. The data sets from Nilsson et al.~\cite{Nilsson:2005} and Shima et al.~\cite{Shima:2005} are chosen here as the latest examples of controversial  experiments characterizing the $^4$He photo-effect since the 50's (see reviews of available data in Refs.~\cite{CBD:1983,quaglioni:044002} and~\cite{Shima:2005}). Note that in the upper panel of Fig.~\ref{csfig2}, we estimate the total cross section from the  $^4$He$(\gamma,n)$ measurements 
of  Ref.~\cite{Nilsson:2005} by assuming $\sigma_{\gamma}(\omega)\simeq2\sigma_{\gamma,n}(\omega)$.  The latter assumption, which relies on the similarity of the $(\gamma,p)$ and $(\gamma,n)$ cross sections, provides a sufficiently safe estimate of the total cross section below the three-body break-up threshold ($\omega=26.1$ MeV).
At higher energies it represents a lower experimental bound for the total cross section, as in the energy range considered here the data of Nillsson et al. do not contain the contributions of the $^4$He$(\gamma,np)d$ and four-body break-up channels.  Shima {\it et al}.~\cite{Shima:2005} provide total photo-disintegration data obtained by simultaneous measurements of all the open channels. Finally, we show also an indirect determination of the photo-absorption cross section deduced from elastic photon-scattering on $^4$He by Wells {et al.}~\cite{Wells:1992}.
We find an overall good agreement with the photo-disintegration data from bremsstrahlung photons~\cite{Nilsson:2005}, which are consistent with the indirect  measurements of Ref.~\cite{Wells:1992}, while we reach only the last of the experimental points of Ref.~\cite{Shima:2005}. 

The lower panel of Fig.~\ref{csfig2} compares our present results with the prediction for the $^4$He photo-absorption cross section obtained in the framework of the EIHH approach~\cite{barnea:054001-03} using the AV18, AV18+UIX~\cite{Gazit:112301} and UCOM~\cite{Bacca:2007} interactions. Interestingly, both the results with AV18 and $\chi$EFT NN interactions and those with AV18+UIX and $\chi$EFT NN+NNN forces show similar peak heights ($\sim 3.2$ mb and $\sim 3.0$ mb respectively), but different peak positions (particularly for the first case) with an overall better agreement of the second set of curves. In this regard we notice that the $\alpha$-particle ground-state properties obtained with AV18+UIX and the $\chi$EFT NN+NNN are very close to each other and to experiment. On the contrary, already at the ground-state level  the two NN interactions are less alike as the $^4$He with the AV18 potential is more than 1 MeV less bound than with the N$^3$LO NN potential, while they still yield to the same point-proton radius.  A somewhat larger discrepancy is found close to threshold between the cross sections obtained with the $\chi$EFT NN+NNN  and UCOM interactions. Beyond $\omega=80$ MeV, in the range not shown in the Figure, the $\chi$EFT NN+NNN force leads to larger cross section values than AV18+UIX or UCOM, which yield to similar results in the region $45\le\omega\le100$ MeV. Keeping in mind that at such high energies the cross section is small and  the uncertainty in our calculation larger, this effect can be related in part to differences in the details and interplay of tensor and spin-orbit forces in the considered interaction models. At the same time corrections to the unretarded dipole operator play here a more important role.

In conclusion we summarize our work. We have calculated the total photo-absorption cross section of $^4$He using the potentials of $\chi$EFT at the orders presently available, the NN at N$^3$LO and the NNN at N$^2$LO. The microscopic treatment of the continuum problem was achieved by means of the LIT method, applied within the NCSM approach. Accurate convergence in the NCSM expansions is reached thanks to the use of three-body effective interactions.  Our result shows a peak around $\omega=27.8$ MeV, with a cross section of $3$ mb.  
The NNN force induces 
a reduction of the peak and an 
enhancement 
of the tail of the cross section. 
The fairly mild NNN effects are far from explaining the low-lying experimental data of Ref.~\cite{Shima:2005} while moderately improve the agreement of the calculated cross section with the measurements of Nilsson {\em et. al.}~\cite{Nilsson:2005}.   In view of the overall good agreement between the $\chi$EFT NN+NNN and AV18+UIX calculations, the photo-absorption cross section at low energy appears to be more sensitive to change in the $\alpha$-particle size, than to the details of the spin-orbit component of the NNN interaction.  In this regard, a more substantial role of the NNN force can be expected in the photo-disintegration of $p$-shell nuclei, for which differences in the spin-orbit strength have crucial effects on the spectrum~\cite{Pieper:2005,Navratil:2007}.
Finally, the rather contained width of the theoretical band embracing the $\chi$EFT NN+NNN, AV18+UIX and UCOM results within 15 MeV from threshold is remarkable compared to the large discrepancies still present among the different experimental data. Hence the urgency for further experimental activity to help clarify the situation.

\section*{Acknowledgments}
We would like to thank Winfried Leidemann for supplying us with the computer code
for the inversion of the LIT. We are also thankful to Giuseppina Orlandini and Sonia Bacca
for useful discussions, and to Ian Thompson for critical reading of the manuscript. 
This work was performed under the auspices of the
U. S. Department of Energy by the University of California, Lawrence
Livermore National Laboratory under contract No. W-7405-Eng-48. Support
from the LDRD contract No.~04--ERD--058 and from
U.S. DOE/SC/NP (Work Proposal Number SCW0498) is acknowledged.




\end{document}